\documentclass[aps,twocolumn,showpacs,amsmath,floatfix]{revtex4}
\usepackage{graphicx}%
\usepackage{dcolumn}
\usepackage{amsmath}
\usepackage{here}

\begin{document}
\newcommand{\be}{\begin{equation}}
\newcommand{\ee}{\end{equation}}
\title{Switching of discrete optical solitons in engineered waveguide arrays}

\author{Rodrigo A. Vicencio$^1$, Mario I.  Molina$^1$, and Yuri S. Kivshar$^2$}

\affiliation{$^1$Departamento de F\'{\i}sica, Facultad de
Ciencias, Universidad de Chile, Casilla 653, Santiago, Chile\\
$^2$ Nonlinear Physics Group and Centre for Ultra-high bandwidth
Devices for Optical Systems (CUDOS), Research School of Physical
Sciences and Engineering, Australian National University, Canberra
ACT 0200, Australia}

\begin{abstract}
We demonstrate simple methods for controlling nonlinear switching
of discrete solitons in arrays of weakly coupled optical
waveguides, for both {\em cubic} and {\em quadratic}
nonlinearities. Based on the effective discrete nonlinear
equations describing the waveguide arrays in the tight-binding
approximation, we develop the concept of the array engineering by
means of a step-like variation of the waveguide coupling. We
demonstrate the digitized switching of a narrow input beam for up
to eleven neighboring waveguides, in the case of the cubic
nonlinearity, and up to ten waveguides, in the case of quadratic
nonlinearity. We discuss our predictions in terms of the physics
of the engineered Peierls-Nabarro (PN) potential experienced by
strongly localized nonlinear modes in a lattice, and calculate the
PN potential for the quadratic nonlinear array for the first time.
We also confirm our concept and major findings for a full-scaled
continuous model and realistic parameters, by means of the beam
propagation method.
\end{abstract}

\pacs{42.82.Et, 42.65.Sf, 42.65.Tg}

 \maketitle

\section{Introduction}

Discrete nonlinear systems are known to support self-localized
modes that exist due to an interplay between a coupling between
the lattice sites and nonlinearity~\cite{phys_today}. Such
spatially localized modes of discrete nonlinear lattices existing
without defects are known as {\em discrete solitons} or {\em
intrinsic localized modes}; they appear in many diverse areas of
physics such as biophysics, nonlinear optics, and solid state
physics~\cite{review,book}. More recently, such modes have been
predicted in the studies of the Bose-Einstein condensates in
optical lattices~\cite{bec} and photonic-crystal waveguides and
circuits~\cite{pbg}.

One of the most important applications of discrete solitons is
found in nonlinear optics where {\em discrete optical solitons}
were first suggested theoretically by Christodoulides and Joseph
~\cite{chrjos88} for an array of weakly coupled optical
waveguides. Because the use of discrete solitons promises an
efficient way to realize and control multi-port nonlinear
switching in systems of many coupled waveguides, this field has
been  explored extensively during last ten years in a number of
theoretical papers (see, e.g.,
Refs.~\cite{kiv93,ace_et96,led_et01}, as an example). More
importantly, the discrete solitons have also been generated
experimentally in fabricated periodic waveguide structures~(see,
e.g., some original papers reporting on the experimental
observations~\cite{eis_et98,meier03} and also the recent review
papers~\cite{silste01,sukh_ieee,chr_nature}).

The majority of theoretical studies conducted so far is devoted to
the analysis of different types of {\em stationary localized
modes} in discrete nonlinear models and their stability.
Consequently, experimental papers have reported on the observation
of self-trapped states in the periodic systems with broken
translational symmetry and some of their properties, in both
focusing and defocusing
regimes~~\cite{silste01,sukh_ieee,chr_nature}. However, only very
few studies and experimental demonstrations addressed more
specific properties of localized modes introduced by discreteness
such as {\em the soliton steering} in and {\em
discreteness-induced trapping} by the lattice (see, e.g.,
Ref.~\cite{dynamics}). As a result, a very little effort has been
made so far to link these findings with realistic applications of
discrete solitons for multi-port all-optical switching.

Indeed, one of the major problems for achieving controllable
multi-port all-optical switching of discrete solitons in waveguide
arrays is the existence of an effective periodic Peierls-Nabarro
(PN) potential which appears due to the lattice discreteness. As a
consequence of this potential, a narrow large-amplitude discrete
soliton does not propagate freely in the lattice and, instead, it
becomes trapped by the array. Several ideas to exploit the
discreteness properties of the array for digitized all-optical
switching have been suggested~\cite{aceves94,bang96}. However, the
main result of those earlier studies is the observation that the
discrete solitons can be well controlled only in the limit of
broad beams whereas the soliton dynamics in highly discrete arrays
has been shown to be more complicated and even
chaotic~\cite{bang96}.

In this paper, we explore in detail an effective way to control
nonlinear switching of discrete solitons in arrays of weakly
coupled optical waveguides earlier suggested in our brief
letter~\cite{ol_ours}.  First, using the discrete model valid in
the tight-binding approximation, we estimate the PN potential
experienced by a strongly localized nonlinear mode that is kicked
initially in a cubic nonlinear waveguide array. The result
suggests a possible control mechanism for the switching of
strongly localized excitations by means of a step-like variation
of the waveguide coupling. For particular types of the engineered
arrays, we are able to demonstrate the digitized switching of a
narrow input beam for up to eleven waveguides. Second, we
demonstrate the validity of the predictions made in the framework
of the discrete model by performing a full-scaled continuous
simulation using realistic parameters. Last but not least, we
extend the concept of controllable digitized switching of discrete
optical solitons to the case of quadratic nonlinear waveguide
arrays, where the experimental observation of discrete optical
solitons has been reported very recently~\cite{chi2_exp}. Here, we
obtain, for the first time to our knowledge, the PN potential for
the discrete soliton and demonstrate numerically the digitized
switching for up to ten waveguides.

The paper is organized as following. In Sec. II we study the
arrays of cubic nonlinear waveguides. First, we consider the
system dynamics described by the discrete nonlinear Schr\"odinger
equation, and show how to modulate the waveguide coupling in order
to suppress the chaotic dynamics and achieve fully controllable
digitized switching. We also employ the beam propagation method
and simulate numerically a more realistic continuous model of the
waveguide arrays with realistic parameters, and confirm that our
concept can be very useful for optimization of the soliton
switching in realistic settings. Next, in Sec. III we extend our
analysis to the arrays of weakly coupled quadratic nonlinear
waveguides, where discrete quadratic solitons are composed of the
coupled beams of the fundamental and second-harmonic fields.
Finally, Sec. IV concludes the paper.

\section{Cubic nonlinear waveguides}

The most common theoretical approach to study the discrete optical
solitons in arrays of weakly coupled optical waveguides is based
on the decomposition of the electric field of the periodic
photonic structure into a sum of weakly coupled fundamental modes
excited in each waveguide of the array;  in solid-state physics
this approach is known as {\em the tight-binding approximation}.
According to this approach, the wave dynamics is described by an
effective discrete nonlinear Schr\"odinger (DNLS) equation that
possesses spatially localized stationary solutions in the form of
discrete localized modes. Many properties of the discrete optical
solitons can be analyzed in the framework of this approach and the
DNLS equation~\cite{chrjos88,led_et01}.

\subsection{Discrete model}
\label{cmt}
\subsubsection{Homogeneous Arrays}

\begin{figure}
\centerline{\includegraphics[width=3.2in]{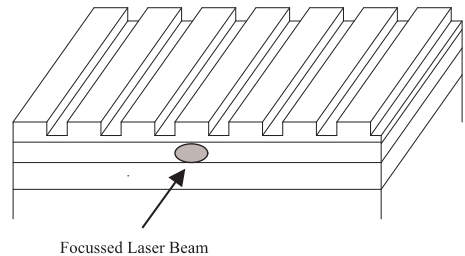}}
\caption{Example of a homogeneous waveguide array and the
generation of a discrete soliton by exciting a single waveguide. }
\label{fig1}
\end{figure}

A standard model of a weakly coupled array of cubic nonlinear
waveguides is described by the DNLS equation~\cite{chrjos88} that
we write in the normalized form \cite{ol_suk_kiv},
\begin{equation}
 i\frac{du_n}{dz} + V ( u_{n+1} + u_{n-1} )
+ \gamma |u_n|^2 u_n = 0, \label{eq:1}
\end{equation}
where $u_n$ is the effective envelope of the electric field in the
$n$-th waveguide, the normalized parameter $V$ is proportional to
the propagation constant of a single waveguide and it
characterizes the coupling between the neighboring waveguides, and
$z$ is the propagation distance along the waveguide. The parameter
$\gamma = \omega_0 n_2/(c A_{\rm eff})$ is the effective waveguide
nonlinearity associated with the Kerr nonlinearity of the core
material. Figure~\ref{fig1} shows a typical experimental structure
of a quasi-one-dimensional homogeneous waveguide array and the
excitation scheme for generating a discrete optical soliton.

Steering and trapping of discrete optical solitons have been
analyzed in the framework of the model~(\ref{eq:1}) in a number of
theoretical studies. Being kicked by an external force, the
discrete soliton propagates through the lattice for some distance,
but then it gets trapped by the lattice due to the discreteness
effects. For a stronger kick, the output soliton position
fluctuates between two (or more) neighboring waveguides making the
switching uncontrollable~\cite{bang96}.

In order to show this feature, first we consider homogeneous
arrays and select an input profile in the form of a narrow
sech-like beam localized on a few waveguides,
\begin{equation}
u_n(0) = A\ {\rm sech} [A(n-n_c)/\sqrt{2}] \
e^{-ik(n-n_c)},\label{eq:sech}
\end{equation}
for $n-n_{c}=0, \pm1$, and $u_n(0)=0$, otherwise. For the particular
results presented below, we consider an array of $101$ waveguides
and place the beam at the middle position, $n_c =50$. The maximum
normalized propagation distance used in our simulations is $z_{\rm
max} =45$. Parameter $k$ in the ansatz (\ref{eq:sech}) has the
meaning of the transverse steering velocity of the beam, in
analogy with the continuous approximation. It describes the value
of an effective kick of the beam in the transversal direction at
the input, in order to achieve the beam motion and shift into one
of the neighboring (or other desired) waveguide outputs.

In our simulations, we control the numerical accuracy by
monitoring the two conserved quantities of model (\ref{eq:1}),
{\em the soliton power}
\begin{equation}
P = \sum_{n} |u_n(z)|^2,
\end{equation}
and {\em the system Hamiltonian},
\begin{equation}
H = - \sum_n \left\{ V (u_n u_{n+1}^* + u_n^* u_{n+1}) +
(\gamma/2) |u_n|^4 \right\}.\label{eq:H}
\end{equation}

\begin{figure}
\centerline{\includegraphics[width=3.75in]{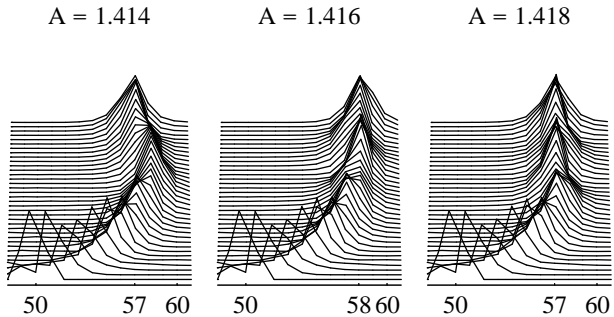}} \caption{An
example of erratic switching of a localized input beam with a
slight variation of the beam intensity in a homogeneous array. }
\label{fig2}
\end{figure}

The input condition (\ref{eq:sech}) does not correspond to an
exact stationary solution of the discrete equation (\ref{eq:1})
even for $k=0$ and, as the input kick ($k \neq 0$) forces the
localized wave move to the right ($k<0$) or left ($k>0$), its
motion is accompanied by some radiation. The effective lattice
discreteness can be attributed to an effective periodic potential,
the PN potential, which is dynamic and changes in time. Due to
both the strong radiation and the presence of the PN barrier which
should be overtaken in order to move the beam transversally, the
discrete soliton gets trapped at one of the waveguides in the
array. In most of the cases, the shift of the beam position to the
neighboring waveguide is easy to achieve, as shown in many studies
\cite{bang96}. However, the soliton switching becomes rather
complicated and even chaotic. This is shown in Fig.~\ref{fig2}
where, for a fixed value of the input angle, a slight variation in
the beam intensity results in a erratic switching of the beam.

\subsubsection{Modulated Arrays}
\label{new}

In this paper, we suggest to modulate the coupling in the
waveguide array in order to achieve a controllable output and to
engineer the switching results. What this modulation of the
couplings does is to affect the PN barrier, providing us with a
simple physical mechanism for fine tuning and control of the beam
self-trapping.


To justify the validity of our concept, we perform a qualitative
estimate of the PN barrier in the framework of the applicability
of the discrete model and perturbation theory. We study the case
of strongly localized modes~\cite{KC} propagating in a homogeneous
waveguide array with identical coupling between the neighboring
waveguides, described by Eq.~(\ref{eq:1}). We consider a general
localized mode that we want to propagate throughout the array. Due
to discreteness, our system lacks the translational invariance
and, as a result, some of its energy must be supplied in order to
force the mode moving. Another way to look at this problem is to
consider that, because of the lattice discreteness, the localized
mode ``sees'' {\em a potential barrier} (the PN barrier), whose
height depends on the effective discreteness of the system as seen
by the excitation~\cite{KC}. Thus, for wide modes, the barrier
will be smaller that for narrow modes. A rough estimate of this PN
barrier can be obtained by equating it to the difference in the
values of the Hamiltonian, between the mode centered at a
waveguide (odd mode) and the mode centered between two neighboring
waveguides (even mode)~\cite{KC}.

In order to evaluate a change of the PN barrier for the mode
initially kicked by an external force, we introduce an initial
phase tilt that is proportional to the factor $\sim \exp (-ikn)$
in the discrete case. Our purpose is not only provide an extension
to the earlier results~\cite{KC}, but also study, for the first
time to our knowledge, the variation of the effective PN potential
for an initially kicked localized mode.

{\em Odd modes.} We consider a strongly localized mode (SLM) in
the form of three excited cites,
\begin{eqnarray}
\lefteqn{u_{n}(z) = u_{n}\ e^{i \lambda_{1} z}}\nonumber\\
                  & \approx &  u_{0}\ \{0..,0,\epsilon_{1} e^{i k}, 1,
\epsilon_{1} e^{-i k}, 0,...0\}\ e^{i \lambda_{1} z},
\label{eq:odd}
\end{eqnarray}
where $u_{0}$ is the mode amplitude, $k$ is the parameter of the
initial ``kick'' (an effective transverse angle) applied to the
mode, $\lambda_{1}$ is the longitudinal propagation constant, and
$\epsilon_{1}$ is a small parameter, to be determined from
Eq.~(\ref{eq:1}). After substituting Eq.~(\ref{eq:odd}) into
Eq.~(\ref{eq:1}) and keeping only linear terms in $\epsilon_{1}$,
we obtain
\[
\lambda_{1} = 2 \epsilon_{1} V \cos(k) + \gamma u_{0}^{2} \]
 and
$\epsilon_{1} = V \cos(k)/\lambda_{1}$, so that
\be \lambda_{1}
\approx \gamma u_{0}^{2},
\hspace{1cm}
\epsilon_{1} \approx V
\cos(k)/\gamma u_{0}^{2} \ll 1.\label{eq:ep1}
\ee

{\em Even modes.} In this case, the SLM mode has the form
\begin{eqnarray}
\lefteqn{ {\tilde u}_{n}(z) = {\tilde u}_{n}\ e^{i \lambda_{2} z} }\nonumber\\
      & \approx &{\tilde u}_{0}\ \{0..,0,\epsilon_{2} e^{i k},
1, e^{-i k}, \epsilon_{2} e^{-2 i k}, 0,...0\}\ e^{i \lambda_{2} z}
\label{eq:even}
\end{eqnarray}
where, as above, ${\tilde u}_{0}$ is the amplitude of the even
mode, $k$ is the initial angle or effective parameter of the
initial ``kick'', $\lambda_{2}$ is the longitudinal propagation
constant of the even mode, and $\epsilon_{2}$ is a small
parameter. After substituting Eq.~(\ref{eq:even}) into
Eq.~(\ref{eq:1}) and keeping only linear terms in $\epsilon_{2}$,
we obtain
\[
\lambda_{2} = (1 + \epsilon_{2})\ V \cos(k) + \gamma {\tilde
u}_{0}^{2} \]
 and $\epsilon_{2} = V \cos(k)/\lambda_{2}$, so that
 \be
 \lambda_{2} \approx V \cos(k) + \gamma {\tilde
u}_{0}^{2}, \epsilon_{2} \approx \frac{V \cos(k)}{V \cos(k) +
\gamma {\tilde u}_{0}^{2}} \ll 1. \label{eq:ep2} \ee

>From Eqs.~(\ref{eq:ep1}) and (\ref{eq:ep2}) we come to the
conclusion that, in order to have strongly localized modes, the
nonlinear contribution described by the term $\gamma u_{0}^{2}$(
or $\gamma {\tilde u}_{0}^{2}$) must be much larger than the
linear term described by the term $V \cos(k)$. Now, for
calculating the PN barrier, we should relate the amplitudes of the
modes of two different symmetries. One way is to think of both the
modes as different states of a single effective mode shifted by a
half lattice site along the chain. This means that the power
content of both modes must be identical, since the power $P =
\sum_{n} |u_{n}(z) |^{2}$ is a conserved quantity. To the first
order in $\epsilon_{1}$ and $\epsilon_{2}$, we obtain
\be P_{\rm odd}  =  u_{0}^{2} + O(\epsilon_{1}^{2}),
\hspace{1cm}P_{\rm even} = 2 {\tilde u}_{0}^{2} +
O(\epsilon_{2}^{2})
\ee
Thus, the relation $P_{\rm odd} = P_{\rm even}$, implies
$u_{0}^{2} \approx 2 {\tilde u}_{0}^{2}$. We are now in position
to compute $H_{\rm odd}$ and $H_{\rm even}$ for a strongly
localized mode, using the above relation and Eqs.~(\ref{eq:H}),
(\ref{eq:ep1}), and (\ref{eq:ep2}),
\begin{eqnarray}
H_{\rm odd} & \approx &- \frac{\gamma}{2}\ u_{0}^{4} + O(\epsilon_{1}^{2})\nonumber\\
H_{\rm even} & \approx & - \frac{\gamma}{4}\ u_{0}^{4} - 2
u_{0}^{2} V \cos(k) + O(\epsilon_{1}\cdot \epsilon_{2}),
\end{eqnarray}
which implies that the PN barrier $\Delta^{(3)}$ for the nonlinear
cubic array is given by \be \Delta^{(3)} =   H_{\rm odd} - H_{\rm
even}
        \approx  -\frac{\gamma}{4}\ u_{0}^{4} + 2 u_{0}^{2} V \cos(k).
\label{eq:PN}
\ee

In comparison with the previously obtained result for the PN
barrier~\cite{KC}, Eq.~(\ref{eq:PN}) adds an extra, albeit small,
term that shows how the PN barrier is modified for the mode
initially kicked in the lattice. Indeed, besides the first term
dependent on the mode amplitude, Eq.~(\ref{eq:PN}) includes a
linear term proportional to the factor $V \cos(k)$, whose
magnitude could be modified by a judicious adjustment of the
waveguide  couplings and/or the value of the initial kick.

Dependence of the PN barrier on the mode coupling suggests that,
if we wish to find a way to engineer the value of the PN barrier
in the lattice, we should study the properties of a modified model
described by the evolution equation
 \be \label{new_eq1}
 i\frac{du_n}{dz} + V_{n+1} u_{n+1} + V_{n-1} u_{n-1}
+ \gamma |u_n|^2 u_n = 0,
\ee
where the coupling $V_{n}$ between two neighboring guides is
assumed to vary either through the effective propagation constant
or by a change in the spacing between the neighboring waveguides.
To study the beam steering in this novel model, we use again as an
initial condition the sech-like profile~(\ref{eq:sech}), although
this is not really fundamental limitation, as argued below.

We mention that a variation of the waveguide coupling in the array
constitutes the starting point for our concept of the waveguide
array engineering. A change of the couplings {\em breaks the
symmetry between the beam motion to the right and left at the
moment of trapping}, thus eliminating chaotic trapping observed in
the case of homogeneous arrays.

\begin{figure}
\centerline{\includegraphics[width=3.75in]{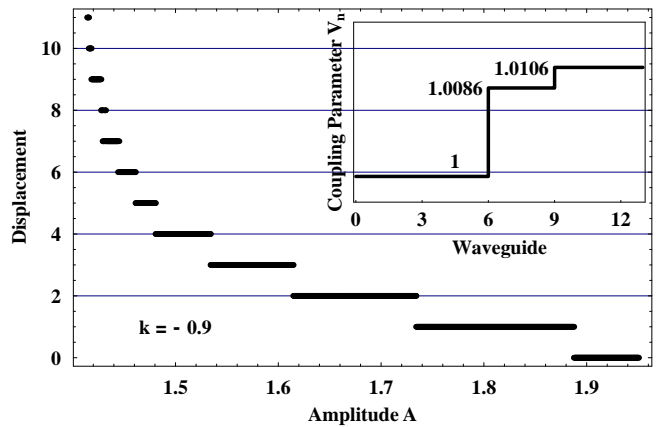}}
\caption{Digitized controlled switching of a discrete soliton in a
cubic nonlinear waveguide array by varying the beam intensity, for
a fixed angle ($k = -0.9$). Inset: Optimized modulation of the
couplings $V_{n}$.} \label{fig3}
\end{figure}

We have tested different types of modulation in the array coupling
and the corresponding structures of the waveguide super-lattices.
An example of one of such optimized structure, where we modulate
the coupling parameter $V_n$ in a step-like manner, is shown in
the inset of Fig.~\ref{fig3} which also shows the discrete
position of the soliton at the output as a function of the
amplitude of the input beam, at a fixed value of the steering
parameter $k = -0.9$. In a remarkable contrast with other studies
(see, e.g., Ref.~\cite{bang96}), the coupling modulation allows to
achieve a controllable digitized switching of discrete optical
solitons in the array with very little or no distortion.

\begin{figure}
\centerline{\includegraphics[width=3.75in]{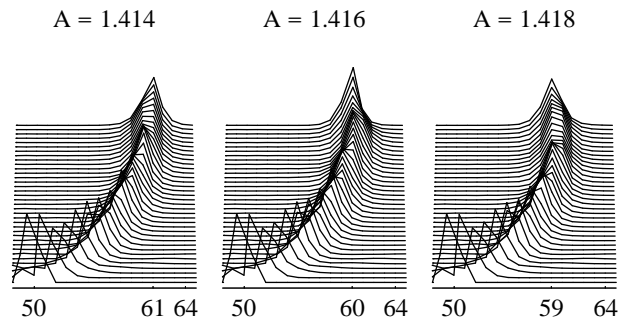}}
\caption{Same as in Fig.~\ref{fig2}, but in the engineered
waveguide array with the coupling modulation shown in the inset of
Fig.~\ref{fig3}.} \label{fig4}
\end{figure}


As is shown in Fig.~\ref{fig4}, by decreasing the amplitude of the
input pulse at a fixed value of the steering parameter $k$ (in our
example fixed to be $k = -0.9$), it is possible to achieve
self-trapping of the discrete soliton by the lattice at some
(short) distance from the input at different waveguide positions.
Due to the step-like modulated coupling, we create a selection
between the beam motion to the right and left at the moment of
trapping thus suppressing or eliminating the chaotic trapping
observed in homogeneous waveguide arrays. In this way, we achieve
{\em a controllable digitized nonlinear switching} where the
continuous change of the amplitude of the input beam results in
{\em a quantized systematic displacement} of the output beam by
{\em an integer number} of waveguides. Consequently, for the
parameters discussed above we observe almost undistorted switching
up to eleven waveguides. Incidentally, we notice here that the use
of a linear ramp potential (e.g., in the form $V_n = a n$) for
this purpose does not lead to an effective switching but, instead,
it makes the soliton switching even more chaotic due to the
phenomenon of Bloch oscillations which become randomized in the
nonlinear regime.

In Fig. 5 we show another example of the optimized coupling
modulation, this time as a function of the effective input `kick',
for a fixed beam intensity. In this case, we can achieve
completely controlled switching up to nine waveguides.

\begin{figure}
\centerline{\includegraphics[width=3.75in]{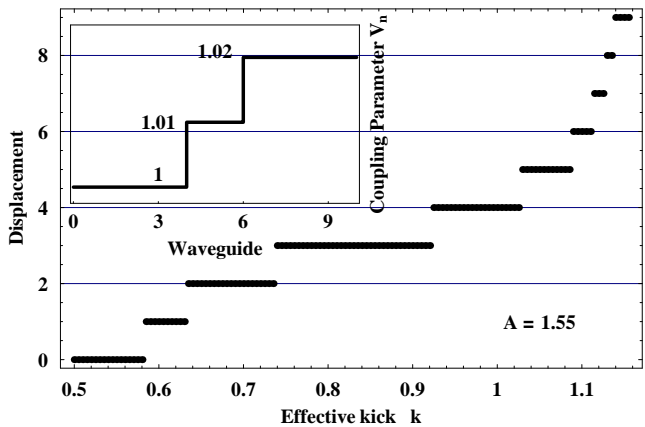}}
\caption{Digitized controlled switching of a cubic discrete
soliton by varying the beam input angle, for a fixed beam
intensity. Inset: Optimized modulation of the couplings $V_{n}$.}
\label{fig5}
\end{figure}

If the input beam was to excite initially five waveguides instead
of three creating in this manner a wider excitation (i.e., being
closer to the continuum limit), one could expect a smaller amount
of radiation emitted. However, this would imply a longer distance
before the beam gets trapped by one of the waveguides in the array
due to the effective PN potential. Also, this means that one
could, in principle, switch the soliton beam to any desired
waveguide in the waveguide array, no matter how far; it would be
just a matter of choosing an initial beam wide enough, i.e.,
closer to the continuum (in addition to optimize the coupling in a
step-wise manner), by removing the random selection between the
directions and suppressing the beam random switching.

Another observation is that the sech-like initial profile is not
really fundamental.  We have verified that the similar dynamics is
observed  for other types of the input beam profiles, including a
nonlinear impurity-like input of the form
\begin{equation}
u_{n}(0) = A \left(\frac{1-A^{2}}{1 + A^2}\right)^{|n-n_{c}|/2}\
e^{-i k (n-n_{c})}.
\end{equation}
The reason for this universal behavior seems to rest on the
observation that for any system with local nonlinearity a narrow
initial profile will render the system into an effective linear
one containing a small nonlinear cluster (or even a single site);
the bound state will therefore strongly resemble that
corresponding to a nonlinear impurity~\cite{last}.

\subsection{Continuous model}

In Sec.~\ref{cmt} we have obtained exceedingly interesting results
for the switching of discrete solitons, {\em v\`{i}a} the use of
the discrete model and tight-binding approximation. In this
section, in order to confirm our predictions, we perform the
corresponding numerical simulations of the continuous evolution
equations of the electric field inside a waveguide array, by means
of the Beam Propagation Method (BPM)~\cite{bpm}. As we show below,
these results support our major findings concerning the digitized
switching of an optical beam in engineered nonlinear waveguide
arrays.

In the continuous model, the starting point is an optical beam
propagating in a three-dimensional medium with the refractive
index that varies in space and is also intensity-dependent
accounting for the Kerr effect. For a nonlinear waveguide array,
the beam is assumed to propagate along the $z$ direction and to
diffract or self-focus along the transversal directions $x$ and
$y$. Assuming that the beam envelope $A(x,y,z)$ varies with $z$ on
a scale much longer that the wavelength $\lambda$, the beam
envelope is found to obey~\cite{yuri_book} the three-dimensional
nonlinear Schr\"{o}dinger (NLS) equation,
\be 2 i \beta_{0} \frac{\partial A}{\partial z} + \left(
\frac{\partial^{2} A}{\partial x^{2}} + \frac{\partial^{2}
A}{\partial y^{2}} \right) + \frac{2 \beta_{0} k_{0} n_{2}}{S}
|A|^{2} A = 0\label{eq:NLS1} \ee
where $\beta_{0} = 2 \pi n_{0}/\lambda$ is the beam propagation
constant, $n_{2}$ is the Kerr coefficient of the nonlinear guide,
$S$ is the area of the mode, and $|A|^{2}$ is the beam power.

The waveguide structure creates a periodic modulation of the
refractive index in only one of the transversal directions (say,
$x$). Along the other transversal direction $y$ we assume the beam
confinement. Therefore, the real dimensionality of the system is
two (i.e., the longitudinal propagation and the transversal
spreading or self-trapping). We write the electric field envelope
in the form $A(x,y,z) = a(x,z) b(y)$ and use the effective index
method~\cite{eim} to formally transform Eq.~(\ref{eq:NLS1}) into
an effective two-dimensional equation
\be i \frac{\partial a}{\partial z} + \frac{1}{2 k_{0} n_{\mbox
{eff}}(x)} \frac{\partial^{2} a}{\partial x^{2}} + \frac{k_{0}
n_{2}}{A_{\mbox {\rm eff}}} |a|^{2} a = 0, \label{eq:NLS2} \ee
where $A_{\mbox {\rm eff}} = S \int_{-\infty}^{\infty} |b|^{2}
dy/\int_{-\infty}^{\infty} |b|^{4} dy$, is the effective area over
which the nonlinear interaction occurs. Parameter $n_{\mbox{eff}}$
is the effective, space-varying {\em linear} index of refraction
for the one-dimensional problem. For our problem,
$n_{\mbox{eff}}(x)$ consists of a periodic array of parallel slabs
with the indices $n_{0}$ and $n_{0} + \Delta n$.

\begin{figure*}
\centerline{\includegraphics[width=7.5in]{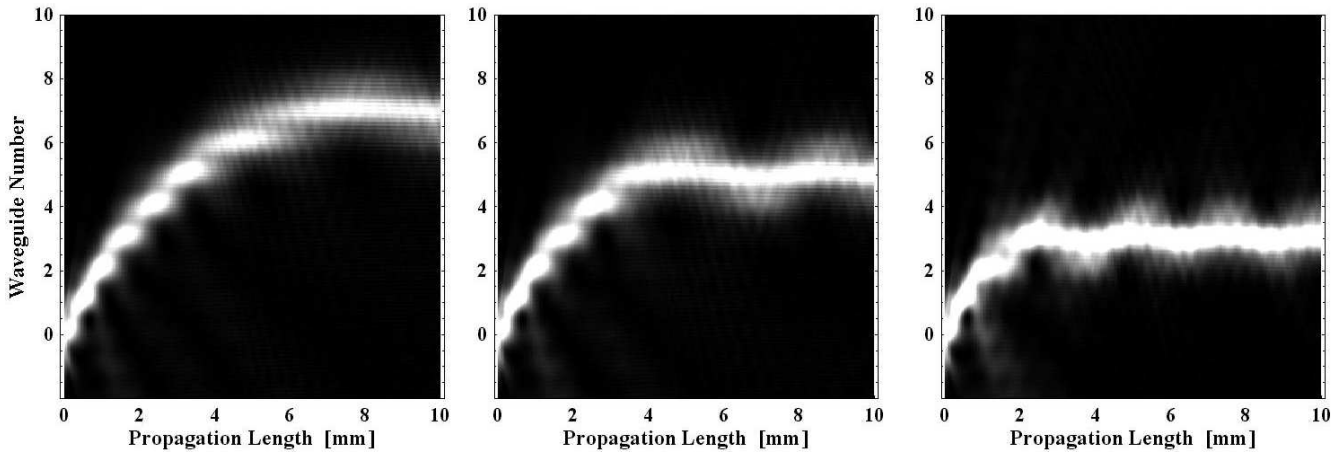}}
\caption{Results of numerical BPM simulations of the continuous
model for the switching of a $1.5 \mu m$ Gaussian beam propagating
in an array of 41 AlGaAs waveguides as a function of the input
beam intensity, for a fixed input beam angle ($0.86^{o}$). Left:
2200 W, center: 2330 W, right: 2700 W.}
\label{fig6}
\end{figure*}

We solve Eq.~(\ref{eq:NLS2}) numerically by the well-known beam
propagation method (BPM), using typical experimental parameter
values. In our simulations, we use an array of 41 waveguides, $10
mm$ long, a Kerr coefficient $n_{2} = 1.5 \times 10^{-17} m^{2}/W$,
assuming a Gaussian beam of the height $3 \mu m$, $8 \mu m $ wide,
at $\lambda = 1.5 \mu m$, and $0.86^{o}$ input angle. The
effective one-dimensional medium consists of a periodic array of
slabs, $4 \mu m$ wide, with $6 \mu m$ center-to-center separation,
$n_{0} = 3.27$, with a modulation of the refractive index $\Delta
n = 0.0014$.

Figure~\ref{fig6} shows the beam switching as the initial input
beam power is varied: Switching to seven, five and three guides
was observed for a power of $2200\ W$, $2330\ W$ and $2700\ W$,
respectively. As expected from the estimates of section
{\ref{new}}, as power is decreased, so does the PN barrier, which
allows the beam to get self-trapped farther away from the vicinity
of the input guide.

The above constitute a strong confirmation of the robustness of
the switching mechanism predicted by a simple theory, against more
realistic effects. Fine tuning of the switching by an appropriate
modulation of the effective waveguide couplings will allow a fine
control of the position of the destination waveguide. However,
this task is beyond the main scope of the present paper.

\section{Quadratic nonlinear waveguides}

Up to now we discussed the arrays of weakly coupled waveguides
with the cubic nonlinearity. However, during last years a growing
interest is observed in the study of nonlinear optical effects
based on the so-called {\em quadratic nonlinearities}. In contrast
to the conventional studies of quadratic nonlinearities where the
main attention is  centered primarily on parametric processes and
the frequency conversion, more recent works are focused on the
phase modulation of the fundamental as well as the second harmonic
waves~\cite{rev_chi2}. This phase modulation accompanies the
familiar amplitude modulation, being the basis of any frequency
conversion, and it may produce the effects which resemble those
known to occur in cubic nonlinear materials. Typical examples are
all-optical switching phenomena in interferometric or coupler
configurations as well as the formation of spatial and temporal
solitons in planar waveguides (see, e.g., Ref.~\cite{rev_chi2} and
the references therein).

Recently, it was demonstrated
theoretically~\cite{chi2_led,chi2_led2,chi2_led3} that arrays of
quadratic nonlinear waveguides represent a convenient system to
verify experimentally many theoretical predictions for the
dynamics of nonlinear latices with cubic nonlinearity. The first
experimental observation of discrete quadratic solitons has been
reported recently by Stegeman and co-authors~\cite{chi2_exp}, who
demonstrated the formation of {\em discrete quadratic solitons} in
periodically poled Lithium Niobate waveguide arrays, excited with
fundamental wave pulses at a wavelength of 1572 nm. These
experimental observations open many perspectives for employing
much larger nonlinearities provided by nonlinear quadratic
materials. In this section, we extend the concept of the
controlled digitized soliton switching discussed above to the case
of quadratic discrete solitons.

\subsection{Discrete Model}

The standard discrete model for an array of weakly coupled
quadratic nonlinear waveguides has the form~\cite{led_et01}:
\begin{eqnarray}
i \frac{d a_{n}}{d z} + V_{a} ( a_{n+1} + a_{n-1}) + 2 \gamma_{2} b_{n} a_{n}^{*} & = & 0\nonumber\\
i \frac{d b_{n}}{d z} + V_{b} ( b_{n+1} + b_{n-1}) + \beta b_{n} +
\gamma_{2} a_{n}^{2} & = & 0,
\label{eq:quad}
\end{eqnarray}
where $a_{n}$ and $b_{n}$ represent the amplitudes for the
fundamental ($\omega$) and second harmonic ($2 \omega$) fields in
the $n$-th guide, $V_{a}$ and $V_{b}$ stand for the linear
couplings between the nearest-neighbor waveguides. Parameter
$\gamma_{2}$ describes the nonlinear second-order coefficient
proportional to the second-order dielectric susceptibility, and
$\beta$ is the effective mismatch between the fields in the array.

As in the case of the cubic nonlinearity, the system
(\ref{eq:quad}) possesses two conserved quantities: {\em the total
power},
\be P = \sum_{n}\left( | a_{n}(z)
|^{2} + 2 | b_{n}(z) |^{2} \right)\label{eq:power_quad}
\ee
and {\em the system Hamiltonian},
\begin{eqnarray}
H & = & -\sum_{n} ( V_{a} a_{n}^{*} a_{n+1} + V_{b} b_{n}^{*} b_{n+1} + (\beta/2) | b_{n} |^{2} + \nonumber\\
  & +  & \gamma_{2} a_{n}^{2} b_{n}^{*} + c.c. ).
  \label{eq:H_quad}
\end{eqnarray}
However, unlike the case of the cubic nonlinear waveguide arrays
where it is possible to find analytical solutions in the continuum
limit which can be used as input profiles for numerical
simulations of discrete solitons, in the case of the quadratic
nonlinearities no exact solutions are available. Thus, we should
resort to the limit of strongly localized modes (SLMs) in order to
calculate the PN barrier and use the SLM profile as an input beam
profile for the numerical computation of the soliton switching.

\subsection{Localized modes and the PN barrier}

As in the case of the cubic nonlinearity, we calculate the PN
barrier as a difference between the values of the Hamiltonian for
the odd and even strongly localized two-component modes.

{\em Odd modes.}  We search for approximate solutions of
Eq.~(\ref{eq:quad}) of the form
\begin{eqnarray}
a_{n} & = & a_{0} \{ ....,0, a_{1} e^{i k},   1, a_{1} e^{-i k}, 0, ...\}\ e^{i \lambda_{1} z} \nonumber\\
b_{n} & = & b_{0} \{ ....,0, b_{1} e^{2 i k}, 1, b_{1} e^{-2 i k},
0, ...\}\ e^{2 i \lambda_{1} z}, \label{eq:quad_odd}
\end{eqnarray}
where $a_{0}$ and $b_{0}$ are the amplitudes of two harmonics
composing a localized mode, $k$ is the initial beam angle or
effective `kick', $\lambda_{1}$ is the longitudinal propagation
constant, and $a_{1}$ and $b_{1}$ are small parameters that should
be determined from the equations. After substituting the ansatz
(\ref{eq:quad_odd}) into Eqs.~(\ref{eq:quad}) and keeping only
linear terms in $a_{1}$ and $b_{1}$, we obtain: $\lambda_{1} = 2
a_{1} V_{a} \cos(k) + 2 \gamma_{2} b_{0}$, $a_{0}^{2} =
(b_{0}/\gamma_{2})[2 \lambda_{1} - \beta - 2 b_{1} V_{b} \cos(2
k)]$, $a_{1} = (V_{a}/\lambda_{1}) \cos(k)$, and $b_{1} = V_{b}
\cos(2 k)/( 2 \lambda_{1} - \beta)$. From these relations, we find
$\lambda_{1} \approx 2 \gamma_{2} b_{0}$, which implies
\begin{eqnarray}
a_{0}^{2} & \approx & 4 b_{0}^{2} - (\beta/\gamma_{2}) b_{0}, \nonumber\\
          &         & a_{1} \approx \frac{V_{a} \cos(k)}{2 \gamma_{2} b_{0}} \ll
          1,\;\;\;
b_{1} \approx \frac{V_{b} \cos(2 k)}{4 \gamma_{2} b_{0} - \beta}
\ll 1.\label{cinco}\ \ \ \ \
\end{eqnarray}

{\em Even modes.}  Now we search for approximate solutions of
Eqs.~(\ref{eq:quad}) of the form
\begin{eqnarray}
\lefteqn{ {\tilde a}_{n}  = {\tilde a}_{0} \{ ..,0,{\tilde a}_{1} e^{i k},1,e^{-i k },{\tilde a}_{1} e^{-2 i k},0, ...\}e^{i \lambda_{2} z} },\nonumber\\
{\tilde b}_{n} & = & {\tilde b}_{0} \{ ..,0,{\tilde b}_{1} e^{2 i
k},1,e^{-2 i k},{\tilde b}_{1} e^{-4 i k},0, ...\}e^{2 i \lambda_{2} z}\ \ \ \ \ \ \ \ 
 \label{eq:quad_even}
\end{eqnarray}
where ${\tilde a}_{0}$ and ${\tilde b}_{0}$ are the amplitudes of
the  coupled harmonics, $k$ is the initial beam angle or effective
`kick', $\lambda_{2}$ is the longitudinal propagation constant,
${\tilde a}_{1}$ and ${\tilde b}_{1}$ are small parameters
determined from the equations of motion. After substituting
Eq.~(\ref{eq:quad_even}) into Eq.~(\ref{eq:quad}) and keeping only
linear terms in ${\tilde a}_{1}$ and ${\tilde b}_{1}$, we obtain:
$\lambda_{2} = (1 + {\tilde a}_{1} ) V_{a} \cos(k) + 2 \gamma_{2}
{\tilde b}_{0}$, ${\tilde a}_{0}^{2} = ({\tilde
b}_{0}/\lambda_{2})( 2 \lambda_{2} - \beta - ( 1 + {\tilde b}_{1}
) V_{b} \cos(2 k))$, ${\tilde a}_{1} = (V_{a}/\lambda_{2})
\cos(k)$ and ${\tilde b}_{1} = V_{b} \cos(2 k)/(2 \lambda_{2} -
\beta)$. From these relations, we find $\lambda_{2} \approx V_{a}
\cos(k) + 2 \gamma_{2} {\tilde b}_{0}$, which implies
\begin{eqnarray}
{\tilde a}_{0}^{2} & \approx & 4 {\tilde b}_{0}^{2} - ({\tilde b}_{0}/\gamma_{2})[\beta - 2 V_{a} \cos(k) + V_{b} \cos(2 k)],\nonumber\\
{\tilde a}_{1}     & \approx & \frac{V_{a} \cos(k)}{V_{a} \cos(k) + 2 \gamma_{2} {\tilde b}_{0}} \ll 1,\nonumber\\
{\tilde b}_{1}     & \approx & \frac{V_{b} \cos(2 k)}{2 V_{a}
\cos(k) + 4 \gamma_{2} {\tilde b}_{0} - \beta} \ll 1.
\label{siete}
\end{eqnarray}

From Eq.~(\ref{cinco}) and Eq.~(\ref{siete}), it is easy to see
that, in order to obtain SLM, the nonlinear term $\gamma_{2} b_{0}
({\tilde b}_{0})$ should be be much larger than the linear
coupling terms, $V_{a} \cos(k)$ and $V_{b} \cos(2 k)$.

From Eq.~(\ref{eq:power_quad}) we calculate the total power
\be P_{\rm odd} \approx a_{0}^{2} + 2 b_{0}^{2} + O(a_{1}^{2},
b_{1}^{2}), \ee
\be P_{\rm even} \approx 2 {\tilde a}_{0}^{2} + 4
{\tilde b}_{0}^{2} + O({\tilde a}_{1}^{2}, {\tilde b}_{1}^{2}),
\ee and the Hamiltonian of each mode,
\begin{eqnarray}
H_{\rm odd} & \approx & -8 a_{0}^{2} a_{1} V_{a} \cos(k) - 8 b_{0}^{2} b_{1} V_{b} \cos(2 k)  \nonumber\\
        &         & - \beta b_{0}^{2} - 2 \gamma_{2} a_{0}^{2} b_{0} + O(a_{1}^{2},
        b_{1}^{2}),
\end{eqnarray}
\begin{eqnarray}
\lefteqn{H_{\rm even} \approx - 4 {\tilde a}_{0}^{2} ( 1 + 2
{\tilde a}_{1} ) V_{a}
\cos(k) - 4 \gamma_{2} {\tilde a}_{0}^{2} {\tilde b}_{0}}\nonumber\\
         &         & -4 {\tilde b}_{0}^{2} ( 1 + 2 {\tilde b}_{1} ) V_{b} \cos(2 k) -
    2 \beta {\tilde b}_{0}^{2}  + O({\tilde a}_{1}^{2}, {\tilde
    b}_{1}^{2}).
\end{eqnarray}

We follow the same reasoning as in the case of the cubic nonlinear
waveguide arrays and calculate the effective PN barrier. Such
calculations look simpler for the physically important case of
vanishing mismatch, $\beta \approx 0$. With that assumption, and
imposing that the power content of both, odd and even modes, are
equal, $P_{\rm odd} = P_{\rm even}$, we obtain $3 b_{0}^{2}
\approx 6 {\tilde b}_{0}^{2} + ( {\tilde b}_{0}/\gamma_{2})(\ 2
V_{a} \cos(k) - V_{b} \cos(2 k)\ )$, and then
\be
{\tilde b}_{0} \approx \frac{b_{0}}{\sqrt{2}} - \frac{(\ 2 V_{a}
\cos(k) - V_{b} \cos(2 k)\ )}{12 \gamma_{2}}.
\ee
In terms of $b_{0}$, the Hamiltonian of both the modes can be
approximated as
\be
H_{\rm odd} \approx -8 \gamma_{2} b_{0}^{3} +
O(a_{1}^{2}, b_{1}^{2})
\ee
\begin{eqnarray}
H_{\rm even} & \approx & -4 \sqrt{2} \gamma_{2} b_{0}^{3} - 8 V_{a} b_{0}^{2} \cos(k) \nonumber\\
         &         & -2 V_{b} b_{0}^{2} \cos(2 k) + O(a_{1}\cdot {\tilde a}_{1}, b_{1}\cdot {\tilde b}_{1}).
\end{eqnarray}
Finally, we calculate, in this approximation, the PN barrier of
the strongly localized modes,
\begin{eqnarray}
\lefteqn{\Delta^{(2)}  =  H_{\rm odd} - H_{\rm even} \approx }\nonumber\\
             &   & -8 C \gamma_{2} b_{0}^{3} + 2 b_{0}^{2}(\ 4 V_{a} \cos(k) + V_{b} \cos(2 k)\ ),
             \label{PN_chi2}
\end{eqnarray}
where $C=( 1 -\sqrt{2}/2 )$.  The PN barrier of an array of
nonlinear quadratic waveguides (\ref{PN_chi2}) has been obtained,
to the best of our knowledge, for the first time. It shows some
interesting features: The main term (\ref{PN_chi2}) is cubic in
the mode amplitude, while for the cubic case it was quartic [see
Eq.~(\ref{eq:PN})]. Also we notice that the first correction to
the PN barrier (\ref{PN_chi2}) is linear in the couplings, and it
depends on the square of the SLM amplitude. This is exactly the
same term as in the case of the nonlinear cubic array. This
implies that the first-order correction is more important in the
nonlinear quadratic array than that in the nonlinear cubic array
suggesting that the appropriate engineering of the couplings
and/or input `kick' to achieve digitized switching should be
easier to achieve.

\begin{figure}
\centerline{\includegraphics[width=3.75in]{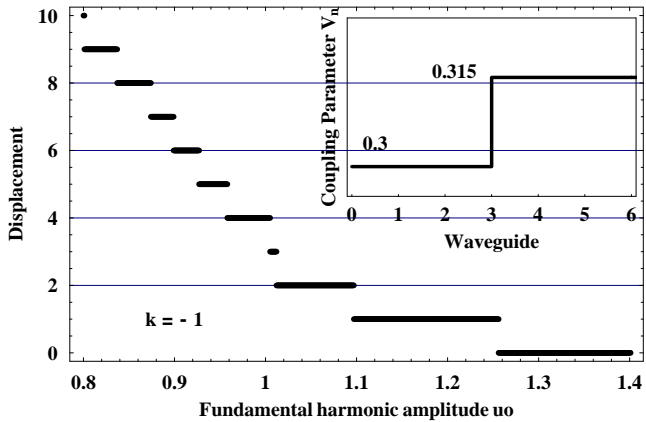}}
\caption{Controlled digitized switching of a discrete quadratic
soliton by a variation of the intensity of the fundamental mode,
for a fixed input `kick'. Inset: Optimized modulation of
$V_{n} = V_{a, n} = V_{b, n}$.} \label{quadratic}
\end{figure}

\begin{figure}
\centerline{\includegraphics[width=3.75in]{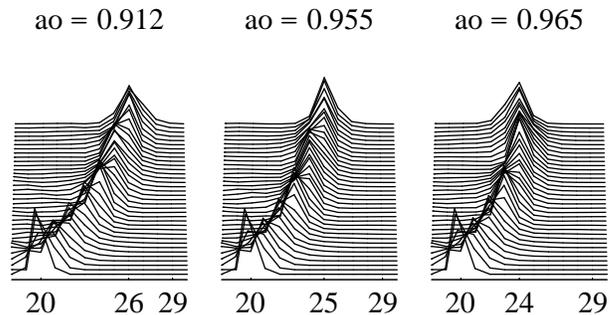}}
\caption{Switching to $6$, $5$ and $4$ sites of a discrete
quadratic soliton (SLM) with a slight intensity variation of the
fundamental mode.} \label{profile_quad}
\end{figure}

For the numerical simulations, we use the initial input in the
form of an odd mode,
\begin{eqnarray}
a_{n}(0) & = & a_{0}\cdot a_{1} ^{|n-n_{c}|} e^{-i(n-n_{c}) k}, \nonumber\\
b_{n}(0) & = & b_{0}\cdot b_{1}^{|n-n_{c}|} e^{-2 i (n-n_{c})
k},
\label{eq:num}
\end{eqnarray}
for $n - n_{c} = 0,\pm 1$, and $a_{n}(0) = b_{n}(0) =0$,
otherwise. In Eq.~(\ref{eq:num}) we use
$a_{0} \approx \sqrt{4 b_{0}^{2}-(\beta/\gamma_{2})b_{0}},
a_{1} \approx V_{a}/2
\gamma_{2} b_{0}$ and $b_{1} \approx V_{b}/(4 \gamma_{2} b_{0} -
\beta)$. We consider an array of 41 waveguides with the initial
input centered at the middle, $n_{c} = 20$. For simplicity, we
also assume the case of complete phase-matching, i.e. $\beta
\approx 0$, and identical coupling for both the harmonic fields,
$V_{a, n} = V_{b, n}$. Figure~\ref{quadratic} shows that the
controlled digitized switching of the discrete two-frequency
(fundamental + second harmonic) soliton can be achieved for up to
10 waveguides, by varying the intensity of the input fundamental
mode for a fixed parameter $k = -1$. The inset of
Fig.~\ref{quadratic} shows the coupling modulation required to
achieve this type of engineered soliton switching, which is
particularly simple and consists of only a single change of about
$5\%$ in the value of the coupling parameter.
Figure~\ref{profile_quad} demonstrates the switching of the
discrete mode, composed of the fundamental and second-harmonic
fields, to six, five, and four neighboring waveguides, as the
intensity of the input fundamental mode is increased. In this
respect, it is interesting to point out that in all cases of the
digital switching both the fundamental and second-harmonic fields
act as a strongly coupled state, and that no `lagging behind' was
observed of any of the modes with respect to the other.

We have performed other simulations with the quadratic nonlinear
array including the cases $V_{b} = 0$ (decoupled second-harmonic
fields in the array) and $V_{b} = \alpha V_{a}$ (reduced coupling
of the second-harmonic fields) with $\alpha < 1$, etc. In all of
those cases, we have observed the digitized switching of the
discrete solitons by engineering the coupling in the array as
discussed above.


\section{Conclusions}

We have suggested and demonstrated numerically a simple but yet
effective method for controlling nonlinear switching of discrete
solitons in arrays of weakly coupled optical waveguides. We have
demonstrated how to achieve the digitized switching of discrete
optical solitons in weakly coupled arrays of cubic and quadratic
nonlinear waveguides described, in the framework of the
tight-binding approximation, by discrete models such as the DNLS
equation with a step-like variation of the waveguide coupling
parameter. Our approach involves a weak step-like modulation of
the coupling strength (or, equivalently, distance between the
waveguides) in the arrays with the period larger than the
waveguide spacing. Such kind of {\em a super-lattice waveguide
structure} allows to modify the trapping properties of the array
due to discreteness as well as engineer the strength of the
effective Peierls-Nabarro potential arising due to the lattice
discreteness. In particular, we have demonstrated the digitized
switching of a narrow input beam for up to eleven waveguides, in
the case of the cubic nonlinear array, and up to ten waveguides,
in the case of quadratic nonlinear array. We have confirmed our
predictions for a full-scaled continuous model and realistic
parameters by employing the beam propagation method.

\section*{Acknowledgements}

Rodrigo Vicencio acknowledges a support from a Conicyt doctoral
fellowship. Mario Molina and Yuri Kivshar acknowledge a support
from the Fondecyt grants 1020139 and 7020139. Yuri Kivshar thanks
the Department of Physics of the University of Chile for a warm
hospitality in Santiago.

\end{document}